\documentclass[twocolumn,superscriptaddress,amsmath,amssymb,aps,longbibliography,prx]{revtex4-2}
\usepackage[colorlinks=true,urlcolor=blue,citecolor=blue,linkcolor=blue]{hyperref}
\usepackage{txfonts}
\usepackage{graphicx}
\usepackage{dsfont}
\usepackage{booktabs}
\usepackage{amsmath}

\usepackage{braket}
\usepackage{dcolumn} 
\usepackage{makecell}

\renewcommand{\vec}[1]{\boldsymbol{#1}}
\newcommand{\Eq}[1]{Eq.~(\ref{#1})}
\newcommand{\Fig}[1]{Fig.~\ref{#1}}
\newcommand{\Alg}[1]{Alg.~\ref{#1}}
\newcommand{\Tr}{\mathrm{Tr}}

\usepackage{algpseudocode} 
\usepackage{algorithm}
\usepackage{varwidth}

\algnewcommand\algorithmicforeach{\textbf{for each}}
\algdef{S}[FOR]{ForEach}[1]{\algorithmicforeach\ #1\ \algorithmicdo}

\begin{document}

\title{Deep Variational Free Energy Approach to Dense Hydrogen}

\author{Hao Xie}
\affiliation{Beijing National Laboratory for Condensed Matter Physics and Institute of Physics, \\Chinese Academy of Sciences, Beijing 100190, China}
\affiliation{School of Physical Sciences, University of Chinese Academy of Sciences, Beijing 100190, China}

\author{Zi-Hang Li}
\affiliation{Beijing National Laboratory for Condensed Matter Physics and Institute of Physics, \\Chinese Academy of Sciences, Beijing 100190, China}
\affiliation{School of Physical Sciences, University of Chinese Academy of Sciences, Beijing 100190, China}

\author{Han Wang}
\email{wang\_han@iapcm.ac.cn}
\affiliation{Laboratory of Computational Physics, Institute of Applied Physics and Computational Mathematics, \\Fenghao East Road 2, Beijing 100094, China}

\author{Linfeng Zhang}
\email{linfeng.zhang.zlf@gmail.com}
\affiliation{DP Technology, Beijing 100080, China}
\affiliation{AI for Science Institute, Beijing 100080, China}

\author{Lei Wang}
\email{wanglei@iphy.ac.cn}
\affiliation{Beijing National Laboratory for Condensed Matter Physics and Institute of Physics, \\Chinese Academy of Sciences, Beijing 100190, China}
\affiliation{Songshan Lake Materials Laboratory, Dongguan, Guangdong 523808, China}

\date{\today}


\begin{abstract}
    We developed a deep generative model-based variational free energy approach to the equations of state of dense hydrogen. We employ a normalizing flow network to model the proton Boltzmann distribution and a fermionic neural network to model the electron wave function at given proton positions. By jointly optimizing the two neural networks we reached a comparable variational free energy to the previous coupled electron-ion Monte Carlo calculation. The predicted equation of state of dense hydrogen under planetary conditions is denser than the findings of {\it ab initio} molecular dynamics calculation and empirical chemical model. Moreover, direct access to the entropy and free energy of dense hydrogen opens new opportunities in planetary modeling and high-pressure physics research.
\end{abstract}

\maketitle 
Hydrogen is the most abundant element in the visible universe. It is also the first and simplest element in the periodic table, consisting of only a proton and an electron. Despite its simplicity, high-pressure dense hydrogen exhibits rich physical phenomena~\cite{McMahon2012} such as metallization~\cite{Wigner1935b} and high-temperature superconductivity~\cite{Ashcroft1968}.  A thorough understanding of these phenomena is of fundamental importance to a broad range of disciplines including planetary physics~\cite{Mazzola2020} and nuclear fusion~\cite{Ichimaru1993}. For these reasons, accurate prediction of the equations of state and phase diagram of dense hydrogen has been a touchstone for computational methods.

Dense hydrogen is a quantum many-body system consisting of coupled protons and electrons. The Fermi temperatures of proton and electron are well separated due to their large mass difference. 
Therefore, for a wide temperature range that falls in between these two scales, one can safely treat protons as classical particles and assume the electrons stay in their instantaneous ground state with fixed proton positions. Solving the electronic Hamiltonian provides an effective potential for the protons. 
However, since the energy scale of electrons is much higher than that of protons, even a tiny error in the electronic calculation could significantly affect the predicted proton configurations. Standard {\it ab initio} molecular dynamics (MD)~\cite{Car1985} solves the electronic structure problem using density functional theory calculations, whose reliability depends on the specific choice of density functionals~\cite{Azadi2013a}. Using machine-learned potential energy surfaces can push such MD simulations to much larger system sizes and longer times~\cite{blank1995neural, jia2020pushing}. However, such an approach at most reflects the accuracy of the underlying electronic structure model that generates the training data, and the reliability issue still persists~\cite{Cheng2020d, Karasiev2021, Zong2020, Tirelli2022, Niu2022}.

A more reliable method to solve the many-electron ground state problem is quantum Monte Carlo~\cite{Foulkes2001}. In the context of dense hydrogen, one can sample the proton configurations according to stochastic estimates of the energy or force acting on the protons, as were previously done in the coupled electron-ion Monte Carlo (CEIMC)~\cite{Pierleoni2004} method and Langevin MD~\cite{Attaccalite2008a}, respectively. However, these nested Monte Carlo approaches have two unsatisfactory drawbacks. First, the statistical noises in the estimated energy or force hamper an unbiased sampling of the protons, similar to the case of Bayesian inference with noisy log-likelihood functions~\cite{Bardenet2017}. There have been three remedies in the literature: a) the noisy Monte Carlo approach~\cite{Kennedy1985} assumes the noises are sufficiently small and treats the acceptance rate using the von Neumann-Ulam method; b) the penalty method~\cite{Ceperley1999} assumes that the noisy energy estimates follow the Gaussian distribution and reduces the acceptance rate with an empirically estimated variance; c) the stochastic gradient Langevin dynamics~\cite{Krajewski2006a,Attaccalite2008a,Welling2011} relies on sufficiently small integration steps with noisy forces to sample from the correct Boltzmann distribution. In all cases, statistical uncertainties in the energy functions deteriorate the sampling efficiency and may even introduce bias to the results. These downsides have partially reduced the reliability advantage of employing more sophisticated quantum Monte Carlo solver for dense hydrogen. For example, suppose one has obtained different results with different flavors of quantum Monte Carlo methods or initial proton configurations, it is rather difficult to tell which results to trust. Second, the nested nature of the approach makes the computation very demanding: one has to make the inner electronic calculation fully converge to ensure correct sampling of the Born-Oppenheimer potential energy surface of protons. Such a stringent requirement makes it rather tedious to ensure convergence of the method~\cite{luo2014ab,Mazzola2017} or limits one to manually crafted variational wave functions with a few or even no variational parameters~\cite{Holzmann2003,Pierleoni2008}. 
 
In light of these difficulties faced by the nested Monte Carlo approaches~\cite{Pierleoni2004,Attaccalite2008a}, we introduce a deep generative model-based variational free energy approach for the dense hydrogen problem. We will minimize the variational free energy with respect to a trial density matrix~\cite{huber1968variational} 
\begin{equation}
F = k_B T \Tr(\rho \ln \rho) + \Tr(\rho H),  
\label{eq:variational-free-energy} 
\end{equation}
with the two terms being entropy and energy respectively; $k_B$ is the Boltzmann constant and $T$ is the temperature. Although free energy minimization is a fundamental principle in quantum statistical mechanics, its practical application is inhibited by the intractable computational cost of entropy term~\footnote{Note that Ref.~\cite{Militzer2000} opted to integrate the Bloch equation for Gaussian density matrix instead of optimizing the variational free energy~\Eq{eq:variational-free-energy}.}. Recent advances in deep generative models~\cite{PILtutorial} have removed this roadblock. Variational free energy calculations based on deep generative models have been applied to a wide range of problems including the Ising models~\cite{Li2018z, Wu2018f, Zhang2018r, PhysRevE.101.023304}, lattice field theories~\cite{Albergo2019,Kanwar2020a,PhysRevLett.126.032001}, atomic solids~\cite{Wirnsberger2021, Ahmad2021a}, quantum dots~\cite{Xie2022a}, and uniform electron gases~\cite{Xie2022}. Because of the intrinsic difficulties of alternative quantum Monte Carlo approaches, the deep generative model-based variational free energy methods have the potential to become an indispensable tool for many-fermion problems at finite temperature, such as Refs.~\cite{Xie2022a, Xie2022} and the hydrogen problem considered here.

\begin{figure}[t]
    \centering
        \includegraphics[width=\columnwidth]{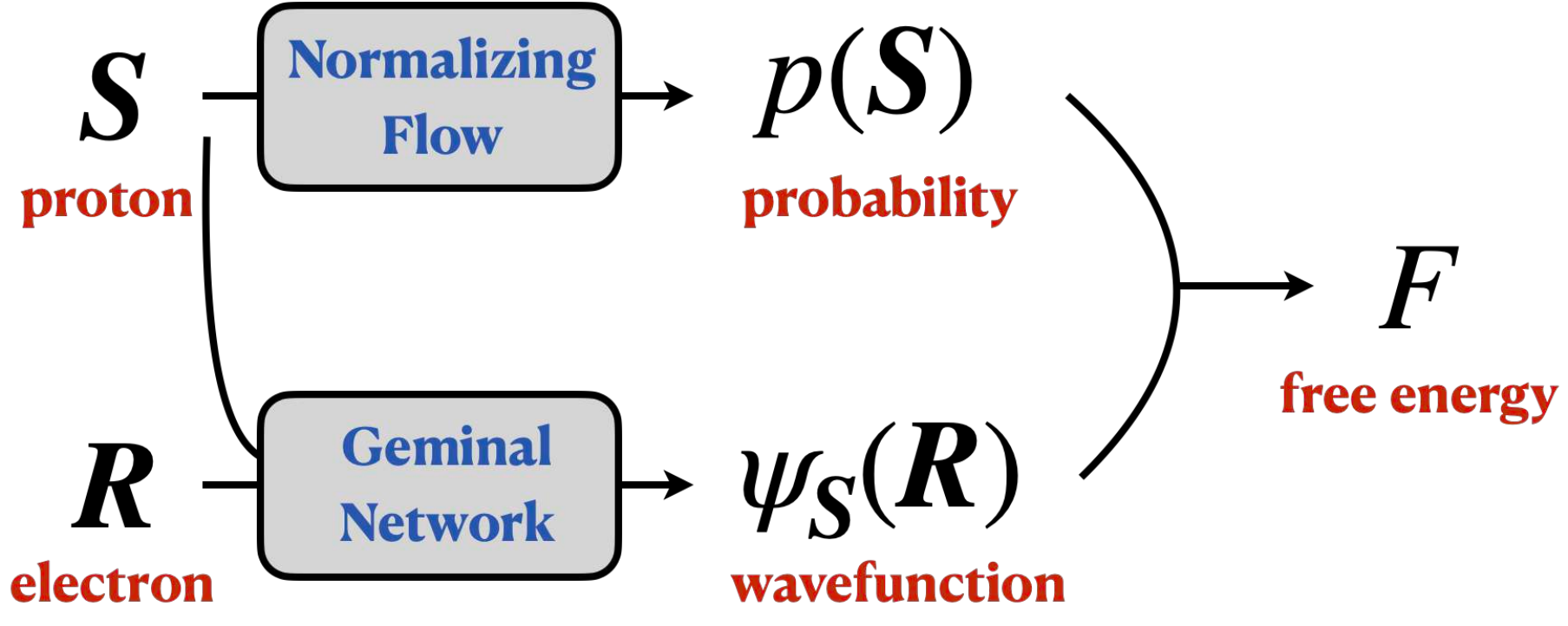}
        \caption{A sketch of computational graph for the dense hydrogen problem. The model consists of a normalizing flow \Eq{eq:flow} for the proton Boltzmann distribution and a geminal neural network~\Eq{eq:wfn} for the electron wave function with fixed proton positions. We jointly optimize the two neural networks to minimize the variational free energy \Eq{eq:variational-free-energy}. 
        \label{fig:concepts}
        }
\end{figure}

We represent the density matrix  $\rho$ using two neural networks as shown in Fig.~\ref{fig:concepts}, one for the proton Boltzmann distribution and one for the electron wave function. Variational free energy calculation of \Eq{eq:variational-free-energy} then amounts to solving a stochastic optimization problem~\cite{Hoffman2012}. In such a formulation, the statistical noises in the estimated energy will not be as catastrophic as in the Monte Carlo sampling~\cite{Pierleoni2004,Attaccalite2008a}. In this respect, the present approach trades the shortcomings of nested Monte Carlo approaches~\cite{Pierleoni2004,Attaccalite2008a} with a variational bias. However, by making use of deep neural network ansatzes one can largely overcome this issue by progressively lowering the variational free energy. With further systematic improvements of variational ansatz and optimization scheme, we anticipate one will reach a reliable description for the whole phase diagram of dense hydrogen with the variational free energy approach~\cite{SM}.

Consider $N$ protons and $N$ electrons in a periodic cubic box of volume $L^3$. The system is unpolarized so there are $N/2$ spin-up and spin-down electrons respectively. The density of the system is specified by the dimensionless parameter $r_s=\left(3/4 \pi N\right)^{1/3}L/a_0$, where $a_0$ is the Bohr radius.
In the atomic units, the Hamiltonian of the hydrogen system reads
\begin{equation}
H =  \sum_i\frac{-\nabla_i^2}{ 2} +  \sum _{i<j} \frac{1}{|\vec{r}_i -\vec{r}_j|}  +  \sum _{I<J} \frac{1}{|\vec{s}_I - \vec{s}_J|} - \sum _{i,I} \frac{1}{|\vec{r}_i - \vec{s}_I|}, 
\label{eq:hamiltonian}
\end{equation}
where $\boldsymbol{S} = \{ \boldsymbol{s}_I \}$ and $ \boldsymbol{R} = \{ \boldsymbol{r}_i \}$ denote the proton and electron coordinates, respectively. Note here we have omitted the proton kinetic energy term, whose effect is considered separately in the Supplemental Material~\cite{SM}. 
 
The variational density matrix of dense hydrogen under consideration is diagonal with respect to the proton degrees of freedom, and can be written as $ \rho  = \int d\vec{S} p(\vec{S}) \Ket{ \vec{S}, \psi_{\vec{S}}} \Bra{ \vec{S}, \psi_{\vec{S}}}$. $ p(\vec{S})  $ is a  normalized probability density for the proton coordinates, and $ \Ket{ \vec{S}, \psi_{\vec{S}}}\equiv \Ket{ \vec{S}} \otimes \ket{\psi_{\vec{S}}} $ is a basis of the whole system's Hilbert space, where $ \Ket{\psi_{\vec{S}}}$ is the electronic ground state at fixed proton configuration $\vec{S}$. The variational free energy $F = \int d \vec{S} \,  p( \vec{S})  \left [ k_B T  \ln p( \vec{S})  +  \Bra{ \psi_{{\vec{S}}  }}  H  \Ket{ \psi_{\vec{S} } }  \right]$ then consists of the entropy of the proton Boltzmann distribution and the electronic expected energy weighted over the proton configurations.

We parametrize the proton Boltzmann distribution $p(\vec{S})$ using a normalizing flow network. Normalizing flow is a class of deep generative model that represents high-dimensional probability density using change-of-variables transformations~\cite{Papamakarios2021}. Specifically, assuming $\vec{\zeta}$ is a set of independent and uniformly distributed ``collective" coordinates, we use a neural network to implement a learnable bijective mapping between $\vec{\zeta}$  and the original proton coordinates $\vec{S}$. The resulting probability density for protons then reads
\begin{equation}
p(\boldsymbol{S}) =  \frac{1}{L^{3N}} \left|\det \left(\frac{\partial \boldsymbol{\zeta}}{ \partial \boldsymbol{S}}\right) \right|.
\label{eq:flow}
\end{equation}
Note this expression is normalized, which facilitates a straightforward and easy computation of the entropy term in the variational free energy objective~\Eq{eq:variational-free-energy}. We construct the bijective transformation between $\vec{\zeta}$ and $\vec{S}$ as a residual network using the fermionic neural network layers~\cite{Pfau2020c}. Since each layer is permutation equivariant, \Eq{eq:flow} is invariant to the permutation of proton coordinates~\cite{Kohlerb}. Moreover, the construction also ensures translational invariance and periodicity of the probability density~\cite{SM}. We compute the transformation Jacobian in \Eq{eq:flow} using forward-mode automatic differentiation~\cite{Baydin2018}. 

Next, we design a neural network for the electronic ground-state trial wave function $\psi_{\vec{S}}(\vec{R}) \propto \Braket{\vec{R} | \psi_{\vec{S}}}$ at fixed proton positions. We concatenate the proton and electron coordinates together and feed them into a fermionic layer~\cite{Pfau2020c} that accounts for the periodic boundary conditions and translational invariance~\cite{Xie2022, PhysRevResearch.4.023138, Wilson2022, Cassella2022, Libe}. The layer outputs $\vec{f}^{\vec{S}} \in \mathbb{R}^{N\times M} $ and  $\vec{f}^\uparrow, \vec{f}^\downarrow \in  \mathbb{R}^{N/2 \times M} $, which are  features that transform equivariantly under permutation of protons or electrons of the same spin. Using these equivariant features, we construct an unnormalized Jastrow-geminal-type wave function~\cite{SM}
\begin{equation}
    \psi_{\vec{S}}(\vec{R})  =  e^J \det \left(G \circ D \right),
\label{eq:wfn}
\end{equation}
where $J = \sum_{i,\mu} a_{\mu} {f}^{\vec{S}}_{i\mu} $ serves as a Jastrow factor, and $\circ$ denotes an element-wise product between two $N/2 \times N/2$ geminal matrices. $G_{ij} =\sum_{\mu\nu} {\chi}^{\uparrow}_{i\mu} W_{\mu\nu} {\chi}^{\downarrow}_{j\nu}$ depends on a learnable real-valued matrix $W$, and $\vec{\chi }^{\uparrow}$, $\vec{\chi}^{ \downarrow} \in \mathbb{R}^{N/2\times M}$ are simply another set of features that are connected with $\vec{f}^{\uparrow}$, $\vec{f}^{\downarrow}$ via a linear map. On the other hand, $D_{ij} 
= \sum_{\vec{k}} \lambda_{\vec{k}} e^{i \vec{k}\cdot (\vec{z}^\uparrow_i-\vec{z}^\downarrow_j)} $ is formed by usual plane-wave orbitals with backflow coordinates $\boldsymbol{z}^{\uparrow}$ and $\boldsymbol{z}^{\downarrow}$. $\lambda_{\boldsymbol{k}}$ are learnable positive parameters representing the occupation number for the momenta $\vec{k}=2\pi \vec{n} / L$ ($\vec{n}\in \mathbb{Z}^3$). We use a large number of $k$ points in the summation so that the wave function can easily capture oscillatory features. 

Note the feature size $M$ relevant to the matrix $G$ plays essentially the same role as the number of plane-wave orbitals in such a geminal-type wave function ansatz. When $M> N/2$, a single geminal determinant would correspond to a summation of combinatorially large $M \choose N/2$ number of determinants according to the Cauchy-Binet formula. From this point of view, the fermionic neural network~\cite{Pfau2020c} construction of the features $\vec{\chi }^{\uparrow}$, $\vec{\chi}^{ \downarrow}$ further boosts the expressibility of the ansatz compared to the well-known geminal states with Jastrow factors~\cite{Becca2017}. 
One can verify that besides being antisymmetric under permutation of electrons of the same spin to account for their fermionic nature, the wave function \Eq{eq:wfn} is also permutation \emph{invariant} with respect to the proton coordinates. 
Note the architecture design here is more lightweight than Refs.~\cite{Gao2021a, Scherbela2021} regarding the goal of respecting the permutation symmetry of nuclei.

By parametrizing the hydrogen density matrix using two neural networks, the variational free energy calculation of \Eq{eq:variational-free-energy} reduces to the following stochastic optimization problem
\begin{equation}
    \min_{\vec{\phi}, \vec{\theta}} \mathop{\mathbb{E}}_{\vec{S}\sim p(\vec{S})} \left[ k_B T \ln p({\vec{S}}) + \mathop{\mathbb{E}}_{\vec{R} \sim |\psi_{\vec{S}}( \vec{R})|^2} \left[  \frac{H\psi_{\boldsymbol{S}}( \boldsymbol{R})}{\psi_{\vec{S}} ( \vec{R}) }\right]  \right], 
\end{equation}
where $\boldsymbol{\phi}, \boldsymbol{\theta}$ are variational parameters of the proton Boltzmann distribution \Eq{eq:flow} and the electronic ground-state wave function \Eq{eq:wfn}, respectively.
We employ the Markov chain Monte Carlo algorithm to draw proton and electron coordinate samples from the two models in an ancestral manner. As for optimization, we have used a generalized stochastic reconfiguration method~\cite{Becca2017} for density matrices, similar to the one employed in~\cite{Xie2022}. Note, however, that some subtle yet important modifications have to be made, which originate from the fact that the electron wave function ansatz~\Eq{eq:wfn} adopted in this Letter is unnormalized~\cite{SM}. 
Since the proton Boltzmann distribution and electron wave function are optimized in a joint manner, one does not need to wait for an expensive inner optimization loop to fully converge before moving the protons. At a conceptual level, jointly moving the nuclear and electronic degrees of freedom is akin to the Car-Parrinello molecular dynamics~\cite{Car1985}, yet in a principled variational optimization framework. 

We employ twist-averaged boundary conditions~\cite{Lin2001} to reduce the finite size effects, in particular those originating from the single-particle momentum shell structure. This amounts to adding some twist angle $\vec{q} \in [-\pi/L, \pi/L]^3$ to the momenta $\vec{k}$ in the plane-wave geminal matrix $D$ of \Eq{eq:wfn}. To further improve the accuracy of electronic calculations, we have also introduced some other explicit twist dependence in the wave function ansatz~\cite{SM}. 

\begin{figure}[t]
    \centering
        \includegraphics[trim={0 0 0 1cm}, width=\columnwidth]{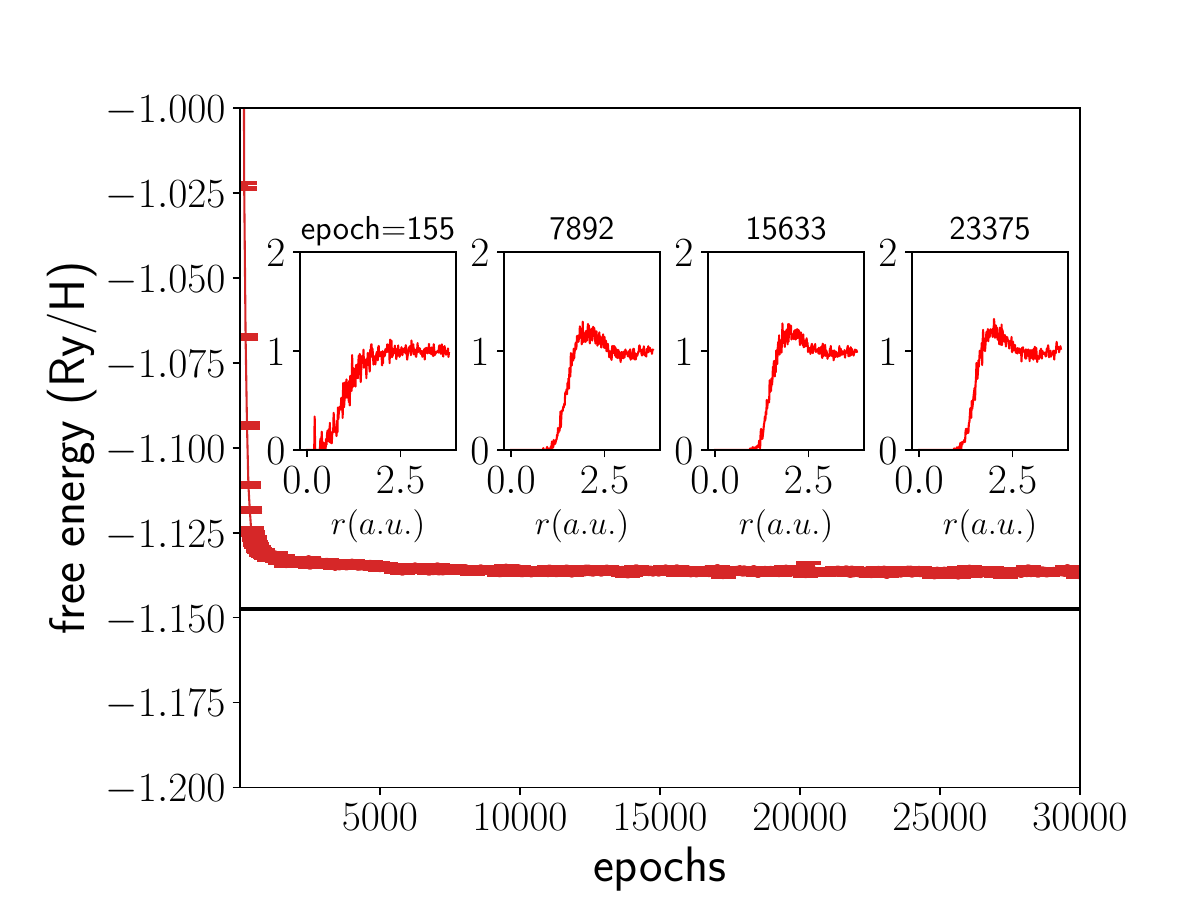}
        \caption{Variational free energy per atom versus optimization epochs of $N=54$ hydrogen atoms at $r_s=1.25$ and $T=6000$K. The horizontal line shows the free energy obtained by coupling-constant integration of CEIMC energies~\cite{Morales2010a}. The inset shows the proton-proton radial distribution functions at several different optimization epochs.
        \label{fig:epochs}}
\end{figure}

As a first application of the deep variational free energy approach, we focus on dense hydrogen at planetary conditions, where it is an ionized liquid of protons and electrons. Equations of state of dense hydrogen under such a condition can be used to construct models for giant planets' interiors~\cite{Mazzola2020}. For these applications, it is also crucial to compute the entropy so that one can follow the adiabatic curve from the planet's surface to its interior~\cite{Miguel2016}. However, standard Monte Carlo methods do not have direct access to the entropy. Reference~\cite{Morales2010a} employed the coupling-constant integration method based on CEIMC calculations to obtain the free energy and entropy of dense hydrogen at $r_s=1.25$ and $T=6000$K . This result was then used as the anchor point to obtain the full equations of state of dense atomic hydrogen. 

We carry out deep variational free energy calculation at the same point ($r_s=1.25$, $T=6000$K) so it is possible to benchmark with Ref.~\cite{Morales2010a}. We choose the system size $N=54$ and perform the twist average over a $4 \times 4 \times 4$ Monkhorst-Pack mesh as in \cite{Morales2010a}, expecting a similar amount of finite size errors as detailed in the Supplemental Material~\cite{SM}. We initialize parameters in the neural networks (\ref{eq:flow}) and (\ref{eq:wfn}) such that the protons are uniformly distributed and the electrons start from an itinerant plane-wave state. Figure~\ref{fig:epochs} shows the variational free energy converges to a slightly higher value than that reported in \cite{Morales2010a}. We note, however, that the reference data may be nonvariational due to the nature of coupling-constant integration. 
The insets of Fig.~\ref{fig:epochs} show the proton-proton radial distribution functions develop a structure as the optimization proceeds. Similar to Ref.~\cite{Morales2010a}, we have found that the proton-proton correlation functions show very little size effect at the considered density and temperature~\cite{SM}.

\begin{figure}[t]
    \centering
        \includegraphics[width=\columnwidth]{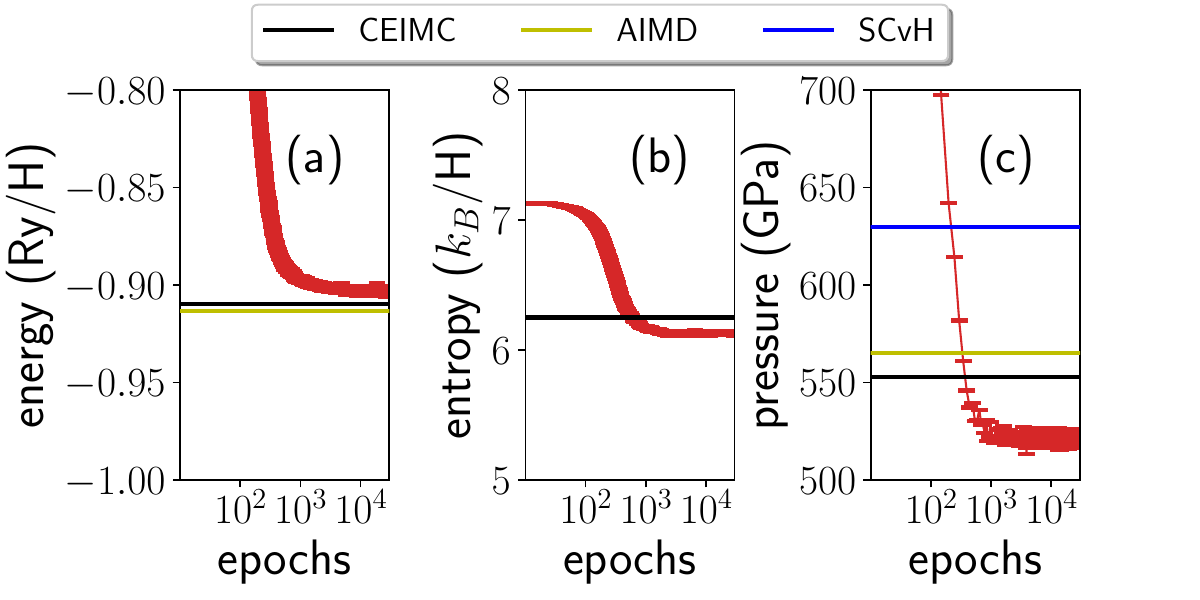}
        \caption{(a) Internal energy per atom, (b) entropy per atom, and (c) pressure versus the optimization epochs. The system parameters are the same as Fig.~\ref{fig:epochs}. The horizontal black lines show the CEIMC results of~\cite{Morales2010a}. The yellow and blue lines show the {\it ab initio} MD and SCvH chemical model~\cite{saumon1995equation} predictions, respectively, also taken from~\cite{Morales2010a}. 
        \label{fig:pressure}}
\end{figure}

Figure~\ref{fig:pressure} shows the energy, entropy (per atom), and pressure of the very same system as Fig.~\ref{fig:epochs}. Notice the entropy decreases upon training, since the protons will develop a more informative distribution than the initially uniform one. The converged values for the energy (entropy) are slightly higher (lower) than the estimates of Ref.~\cite{Morales2010a}, respectively, even after accounting for the finite size corrections~\cite{SM}.
The pressure is computed using the virial theorem as $ (2K+V)/(3 L^3)$~\cite{landau2013statistical}, where $K$ and $V$ are the total kinetic and potential energy of the system, respectively. 

After accounting for the finite size correction~\cite{SM}, our estimated pressure $529(4)$ GPa is slightly smaller than  the CEIMC prediction $553(1)$ GPa~\cite{Morales2010a}. Nevertheless, both values are smaller than {\it ab initio} MD calculation with local density approximation, which is even further away from the Saumon–Chabrier–van Horn (SCvH) equations of state based on a chemical model as a mixture of hydrogen molecules, atoms, protons, and electrons~\cite{saumon1995equation}. Predicting an even denser equation of state than the SCvH chemical model is in line with previous quantum Monte Carlo calculations~\cite{Morales2010a, Mazzola2018}. Combined with the ability to directly access entropy, the deep variational free energy approach developed in this Letter can be a valuable tool for planetary modeling.


The fact that we have reached higher internal energy ($K+V$) but lower pressure ($\propto 2K+V$) indicates our calculation obtains lower kinetic energy and higher potential energy than the CEIMC results of Ref.~\cite{Morales2010a}. It is known that the accuracy of internal energy is often greater than pressure, as the latter requires high accuracy both in the kinetic and potential energies whose errors tend to cancel out in the internal energy~\cite{Militzer2001b}. 
Likewise, we are thus more confident about our calculated internal energy than the pressure because the virial estimator for the pressure 
omits Pulay-like corrections due to incomplete optimizations~\cite{Mazzola2014}. 
We have released our codes and trained models~\cite{github}. By inspecting and further reducing the variational errors one can verify our findings reported here. 


Unlike many previous studies on dense hydrogen~\cite{Pierleoni2004, Attaccalite2008a, Morales2010a, Pierleoni2016, Mazzola2018}, our calculation starts from a rather uninformative point with minimal physical constraints. In this regard, it is rather satisfying that the present calculation has yielded compatible equations of state for such an intensively studied system. In the future, one can also put prior knowledge such as the empirical or machine-learned potential~\cite{DPMD} into the flow model~\Eq{eq:flow}. One can use them either to pretrain the flow model or replace the uniform base distribution. In the latter case, the flow transformation is intended only to learn a small correction and our variational objective function will be the free energy difference with the base model.   

The results of Figs.~\ref{fig:epochs} and \ref{fig:pressure} run on $32$ Nvidia A100 GPUs for about 200 hours. It will just be a matter of parameter scan to produce the full equations of state of dense hydrogen in the atomic phase with direct access to entropy and free energy data. 
For higher temperatures, one can take into account the thermal effect of electrons using the neural canonical transformation approach~\cite{Xie2022a,Xie2022}. Note that unlike the path integral Monte Carlo methods~\cite{cepeiley1996path, PhysRevLett.73.2145, PhysRevE.63.066404, PhysRevLett.129.066402} such an approach will not suffer from the fermion sign problem. At lower temperatures, intriguing physics such as quantum liquid and superconducting order may come into play. To take into account nuclear quantum effects in this case, we envision one can either generalize the proton probabilistic model to the path integral representation~\cite{feynman2010quantum} or adopt the neural canonical transformation approach~\cite{Xie2022a,Xie2022} for protons. Either way, we anticipate the computational cost brought by the zero-point motion of protons does not increase significantly, in contrast to conventional approaches like path integral MD. 

In this Letter, we have directly computed the Jacobian of the flow transformation in \Eq{eq:flow}. To scale up to even larger system sizes, one may employ more efficient permutation equivariant flow models~\cite{Li2020m, Bilos2020, Wirnsberger2020, Wirnsberger2021}. Moreover, one may also consider more scalable optimization schemes~\cite{Martens2015} for a larger number of variational parameters.  In the meantime, it may be useful to explore optimization schemes beyond the score function gradient estimator~\cite{Mohamed2020} for the flow model. For example, one can use the pathwise gradient estimator to exploit information about the nuclear force~\cite{Sorella2010a, Qian2022a}. It is unclear whether this alternative choice will make training of the flow model more efficient. 
Last but not least, in conjunction with an alternative optimization approach, one could also explore the possibility of direct sampling of proton configurations with the normalizing flow. 

It is believed there is a first-order atomic-to-molecular transition in the phase diagram of hydrogen~\cite{McMahon2012}. Near the phase transition, there are significant difficulties related to slow equilibrium time or even lack of ergodicity in the Monte Carlo sampling of multimodal probability distributions~\cite{Mazzola2017}. This difficulty does not magically disappear in the variational free energy approach as they will show up as metastable states in the optimization landscape. Fortunately, the variational nature of our approach provides a clear guidance to judge and improve the qualities of various calculations, as one can always choose to believe the solution with the lowest free energy. We regard this as the most appealing feature of the present approach over the previous ones based on nested Monte Carlo sampling~\cite{Pierleoni2004, Attaccalite2008a}. 

Deep variational free energy optimization presented in this Letter is a general computational framework. Both the nuclear probability distribution and the electronic wave function are open to further extensions. Thus, the framework holds the promise to be applied to a broad range of finite-temperature quantum matters beyond dense hydrogen. 

\begin{acknowledgments}
We thank Xinyu Li, Qi Yang, and Xing-Yu Zhang for their support of computational resources. We thank Guglielmo Mazzola, Mohan Chen, Xinguo Ren, and Quansheng Wu for useful discussions. This project is supported by the Strategic Priority Research Program of Chinese Academy of Sciences under Grants No. XDB0500000 and No. XDB30000000, and National Natural Science Foundation of China under Grants No. 92270107, No. 12188101, No. 12122103, No. T2225018, and No. T2121001. This work is also supported in part by Huawei CSTT Project ``Learning Neural Physics Engines."
\end{acknowledgments}
 
\bibliography{refs,manual_refs}

\setcounter{table}{0}
\renewcommand{\thetable}{S\arabic{table}}
\setcounter{figure}{0}
\renewcommand{\thefigure}{S\arabic{figure}}
\setcounter{equation}{0}
\renewcommand{\theequation}{S\arabic{equation}}

\appendix 

\clearpage 

\section*{Supplemental Material}

\section{Proton kinetic energy contributions to the thermodynamic quantities}

In the Born-Oppenheimer approximation, the proton kinetic energy term is omitted in the Hamiltonian of hydrogen system. As a result, the variational density matrix is diagonal with respect to the proton coordinates $\vec{S}$, and the variational free energy defined in \Eq{eq:variational-free-energy} of the main text satisfies $F \geq -k_B T \ln z$, where $z = \int d\vec{S} e^{- E(\vec{S}) / k_B T}$ is the partition function, and $E(\vec{S})$ is the electronic ground-state energy of the Hamiltonian at given proton configuration $\vec{S}$.

Once the variational free energy minimization is performed, we have to account for extra contributions of the proton kinetic energy term by considering the \emph{full} partition function of the system:
\begin{align}
    Z &= \frac{1}{N! h^{3N}} \int d\vec{P} \int d\vec{S} \exp\left[ -\frac{1}{k_B T} \left( \frac{\vec{P} ^2}{2 m_\textrm{p}} + E(\vec{S}) \right) \right] \nonumber \\
    &= \frac{z}{N! \lambda^{3N}}, 
    \label{eq: full partition function}
\end{align}
where $\vec{P}$ is the proton momenta and $\lambda = h / \sqrt{2 \pi m_\textrm{p} k_B T}$ is the thermal de Broglie wavelength of protons. $h$ is the Planck constant and $m_\textrm{p}$ is the proton mass. The additional prefactor $1/N! \lambda^{3N}$ in \Eq{eq: full partition function} contributes to various thermodynamic quantities. Specifically, it gives an offset $ k_BT \ln ( N! \lambda^{3N})$ to the free energy, $\frac{3}{2}Nk_B T$ to the internal energy, $ \frac{N k_B T/\mathrm{Ry}}{(L/a_0)^3} \frac{\mathrm{Ry}}{a_0^3} =  \frac{3 k_B T/\mathrm{Ry}}{4\pi r_s^3} 14710.5 \mathrm{GPa}$ to the pressure, and $\left[\frac{3}{2}N  - \ln (N! \lambda^{3N}) \right] k_B$ to the entropy. These contributions have been taken into account in all numerical results of this work relevant to finite-temperature calculations.

\section{Model architectures \label{sec:model architecture}}
In this section, we describe in detail the two neural networks that are used to construct the variational density matrix of dense hydrogen. Learnable parameters or layers are indicated in {\color{blue} blue}. 

\subsection{FermiNet layer}
We define a generic \texttt{FermiNet} layer~\cite{Pfau2020c}, shown in Algorithm \ref{alg:ferminet}, as a building block for the modeling of both proton Boltzmann distribution and electron wave function. This network maps a set of particle coordinates $\vec{x}_1, \dots, \vec{x}_D$ to the same number of output features $\vec{f}_1, \dots, \vec{f}_D$ in a translation-invariant and permutation-equivariant way. 

\begin{algorithm}[H]
	\begin{algorithmic}[1]
        \caption{A \texttt{FermiNet} layer for particles in a periodic box.}
        \Require Coordinates $\boldsymbol{x}$, partition indices $\texttt{idx}$, and box length $L$. 
        \Ensure A list of translation-invariant and permutation-equivariant features.  
		\State $\boldsymbol{h_1} = \texttt{zeros\_like}(\boldsymbol{x})$ \Comment single-particle features
	    \State $\boldsymbol{x}_{ij} = \boldsymbol{x}_i - \boldsymbol{x}_j$ 
        \State $\boldsymbol{h_2} = \left[\left| \sin\left(\frac{\pi \vec{x}_{ij}}{L}\right) \right|, \cos\left( \frac{2\pi \boldsymbol{x}_{ij}}{L}\right), \sin\left( \frac{2\pi\boldsymbol{x}_{ij}}{L}\right) \right]$  \Comment two-particle features
		\For{$\ell=1,\cdots ,d$} 
        \State
            \begin{varwidth}[t]{\linewidth}
                $\boldsymbol{g} = [\boldsymbol{h_1},$~\par
                \hskip\algorithmicindent $\texttt{mean}(\boldsymbol{h}, \texttt{axis=0})~\texttt{for}~\boldsymbol{h}~\texttt{in}~\texttt{split}(\boldsymbol{h}_1, \texttt{idx}, \texttt{axis=0}),$~\par
                \hskip\algorithmicindent $\texttt{mean}(\boldsymbol{h}, \texttt{axis=0})~\texttt{for}~\boldsymbol{h}~\texttt{in}~\texttt{split}(\boldsymbol{h}_2, \texttt{idx}, \texttt{axis=0})]$
            \end{varwidth}
		\If {$ \ell = 1$}
		   	\State $\boldsymbol{h_1} = \tanh ({\color{blue}\texttt{FC}_{n_1}^\ell}(\boldsymbol{g}))$   \Comment  $\mathbb{R}^{D \times n_1}  $ 
			\State $\boldsymbol{h_2} = \tanh ({\color{blue}\texttt{FC}_{n_2}^\ell}(\boldsymbol{h_2}))$  \Comment  $\mathbb{R}^{D \times D \times n_2} $ 
		\Else
		    \State $\boldsymbol{h_1} = \tanh ({\color{blue} \texttt{FC}_{n_1}^\ell}(\boldsymbol{g})) +\boldsymbol{h_1} $   
			\State $\boldsymbol{h_2} = \tanh ({\color{blue} \texttt{FC}_{n_2}^\ell}(\boldsymbol{h_2})) + \boldsymbol{h_2} $
		\EndIf 
		\EndFor
        \State
            \begin{varwidth}[t]{\linewidth}
                $\boldsymbol{g} = [\boldsymbol{h_1},$~\par
                \hskip\algorithmicindent $\texttt{mean}(\boldsymbol{h}, \texttt{axis=0})~\texttt{for}~\boldsymbol{h}~\texttt{in}~\texttt{split}(\boldsymbol{h}_1, \texttt{idx}, \texttt{axis=0}),$~\par
                \hskip\algorithmicindent $\texttt{mean}(\boldsymbol{h}, \texttt{axis=0})~\texttt{for}~\boldsymbol{h}~\texttt{in}~\texttt{split}(\boldsymbol{h}_2, \texttt{idx}, \texttt{axis=0})]$
            \end{varwidth}
		\State   $\vec {f} = \tanh ({\color{blue}\texttt{FC}_{n_1}}( \boldsymbol{g}  )) + \boldsymbol{h_1}$  \Comment  $\mathbb{R}^{D\times n_1}$ 
		\State \Return  $\texttt{split}(\boldsymbol{f}, \texttt{idx}, \texttt{axis=0})$ 
		\label{alg:ferminet}
	\end{algorithmic}
\end{algorithm}

To handle an array of concatenated coordinates of protons and electrons of both spins, we have introduced a split operation $\texttt{split}(\vec{x}_{1:D}, \texttt{idx}) := (\vec{x}_{1:\texttt{idx(1)}}, \vec{x}_{\texttt{idx}(1)+1:\texttt{idx}(2)}, \dots, \vec{x}_{\texttt{idx}(K)+1:D})$, where $\texttt{idx}$ is an integer sequence of length $K$ specifying the cutting points of the array. This operation can be straightforwardly generalized to higher-dimensional arrays for any given axis. $\texttt{FC}_n$ denotes a fully connected layer with $n$ output features. In our calculations, the network depth is set to be $d=3$ for the proton distribution and $d=4$ for the electron wave function, and the one- and two-particle feature size are $n_1=32$ and $n_2=16$, respectively. The output feature size is equal to $n_1=32$, which we also refer to as $M$ in the algorithm to be consistent with the main text.

\subsection{Probabilistic model for protons}
The probabilistic model for proton Boltzmann distribution is a normalizing flow with a uniform base distribution, as shown in Algorithm \ref{alg:flow}. Since it only involves $N$ proton coordinates $\vec{S} = \{\vec{s}_1, \dots, \vec{s}_N\}$, no splitting is needed, and the partition index list $\texttt{idx}$ is empty.

\begin{algorithm}[H]
	\begin{algorithmic}[1]
		\caption{Normalizing flow for proton Boltzmann distribution.}
        \Require Proton coordinates $\vec{S}$ and box length $L$. 
        \Ensure Log-probability $\ln p(\vec{S})$
        \State $\boldsymbol{f} ={\color{blue}  \texttt{FermiNet}} (\boldsymbol{S}, [], L)$   \Comment $\mathbb{R}^{N\times M }  $ 
        \State $\boldsymbol{\zeta} = \boldsymbol{S} + {\color{blue}\texttt{FC}_{3}}(\boldsymbol{f})$ \Comment $\mathbb{R}^{N\times 3}$ 
		\State \Return $ \ln \left|\det
		\left( \frac{\partial \boldsymbol{\zeta} }{\partial \boldsymbol{S}} \right )\right|  -\ln ( L^{3N})$
		\label{alg:flow}
	\end{algorithmic}
\end{algorithm}

\subsection{Electron wave function ansatz}
The electron ground-state wave function depends on both the proton coordinates $\vec{S}$ and the electron coordinates $\vec{R} = \{\vec{r}^\uparrow_1, \dots, \vec{r}^\uparrow_{N/2}, \vec{r}^\downarrow_1, \dots, \vec{r}^\downarrow_{N/2}\}$ of both spins. We thus set the partition index list to be $[N, N+N/2]$ to separate the three coordinate sets, and feed the output features $\boldsymbol{f}^{\vec{S}}, \boldsymbol{f}^\uparrow, \boldsymbol{f}^\downarrow$ of the \texttt{FermiNet} layer to a geminal network shown in Algorithm \ref{alg:geminal}. See the main text for detailed discussions about the network design. The resulting wave function is invariant to the permutation of protons, and antisymmetric to the permutation of electrons with the same spin. 
\begin{algorithm}[H]
	\begin{algorithmic}[1]
		\caption{Geminal network for the electron ground-state wave function.}
        \Require Proton coordinates $\boldsymbol{S}$, electron coordinates $\boldsymbol{R} = [ \boldsymbol{r}^\uparrow, \boldsymbol{r}^\downarrow ]$, and box length $L$.
        \Ensure Log-wavefunction $\ln \psi_{\vec{S}}(\vec{R})$. 
        \State $\boldsymbol{f}^{\vec{S}}, \boldsymbol{f}^\uparrow, \boldsymbol{f}^\downarrow = {\color{blue}\texttt{FermiNet}} ([\boldsymbol{S} , \boldsymbol{r}^\uparrow,  \boldsymbol{r}^\downarrow], [N, 3N/2], L)$ 
	    \State $J = \sum_{i,\mu} {\color{blue} a_\mu} f_{i\mu}^{\vec{S}}$ \Comment Jastrow factor
	    \State $\boldsymbol{\chi}^{\uparrow, \downarrow} = {\color{blue}\texttt{FC}_M}(\boldsymbol{f}^{\uparrow, \downarrow})$  \Comment $\mathbb{R}^{N/2 \times M}$
	    \State $G_{ij} = \sum_{\mu\nu} {\chi}^\uparrow_{i\mu} {\color{blue} W_{\mu\nu}}  {\chi}^\downarrow_{j\nu}$   \Comment $\mathbb{R}^{N/2 \times N/2}$
	    \State $\boldsymbol{z}^{\uparrow, \downarrow} = \boldsymbol{r}^{\uparrow, \downarrow} + {\color{blue}\texttt{FC}_{3}}(\boldsymbol{f}^{\uparrow, \downarrow} ) $  \Comment Backflow coordinates  
	    \State $D_{ij} = \sum_{\boldsymbol{k}} {\color{blue} \lambda_{\boldsymbol{k}} } e^{i \boldsymbol{k} \cdot (\boldsymbol{z}^{\uparrow}_i - \boldsymbol{z}^{\downarrow}_j ) }$  \Comment $\mathbb{C}^{N/2 \times N/2}$
	    \State \Return $ J + \ln \det (G \circ D)$ 
		\label{alg:geminal}
	\end{algorithmic}
\end{algorithm}

\begin{figure}[H]
    \centering
        \includegraphics[width=\columnwidth]{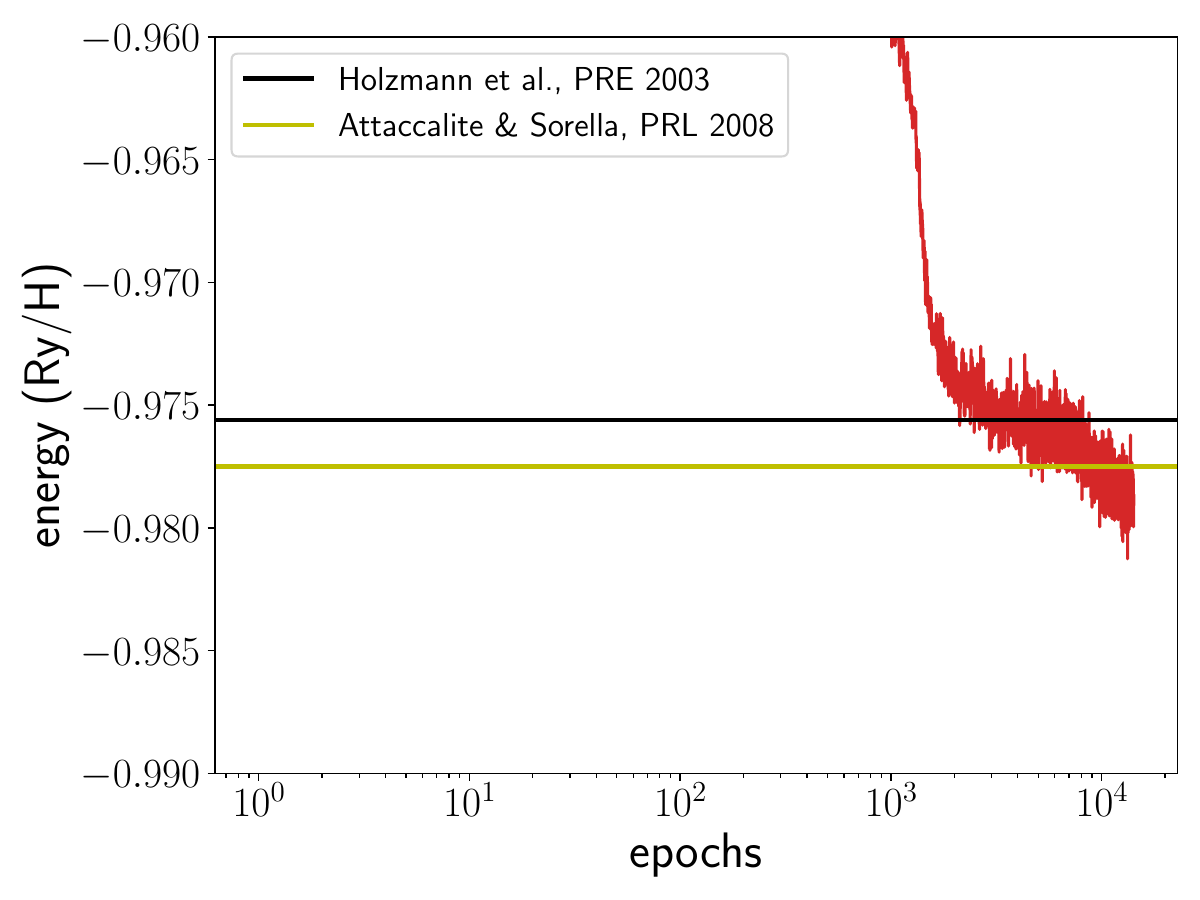}
        \caption{The ground-state energy benchmark of $N=16$ hydrogen atoms arranged in a BCC lattice at $r_s=1.31$. The red curve shows a running average over the optimization epochs, while the horizontal lines are reference data from Refs.~\cite{Holzmann2003} and \cite{Attaccalite2008a}. 
        \label{fig:gs}
        }
\end{figure}

To get a feeling for its accuracy, we use the geminal network to carry out a pure ground-state variational Monte Carlo calculation of $N=16$ hydrogen atoms fixed in a BCC lattice at $r_s = 1.31$. As shown in Figure~\ref{fig:gs}, the optimized energy is lower than previous results, which used conventional (but efficient) wave function ansatzes based on carefully chosen form of Jastrow or backflow correlations~\cite{Holzmann2003, Attaccalite2008a}.

In practice, we initialize the geminal matrix $G$ in \Alg{alg:geminal} to be nearly identity (with respect to the element-wise product ``$\circ$" operation; that is, a matrix of all ones). This can make the neural network close to a usual geminal state with plane-wave orbitals, which is clearly a reasonable initialization for metallic hydrogen at high densities, as demonstrated in the main text. However, due to the (presumably) universal approximating nature of the neural network, there are no \textit{a priori} reasons to rule out the possibility to capture physics of the system in other phase regions. See Appendix~\ref{sec:molecular phase} for an example.

\subsection{Twist-averaged boundary conditions \label{sec:tabc}}
We employ twist-averaged boundary conditions (TABC) in electronic calculations to reduce the finite size effect~\cite{Lin2001}. To do this, we replace the momenta $\vec{k}$ by $\vec{k} + \vec{q}$ in the geminal wave function ansatz shown in the $6$th line of \Alg{alg:geminal}. Note that when wrapping around the periodic box, the spin-up and spin-down electrons will pick up opposite phases~\cite{Mazzola2018}, which reflects the time-reversal symmetry of the system.

Besides the simple replacement $\vec{k} \rightarrow \vec{k} + \vec{q}$, we also introduce some other dependence on the twist $\vec{q}$ in the wave function ansatz to enhance its expressive power. (Not shown in \Alg{alg:ferminet} and \Alg{alg:geminal} above for simplicity.) Firstly, we make the single-particle features $\vec{h_1}$ in the \texttt{FermiNet} layer depend on the twist. Secondly, $\lambda_{\vec{k}}$ are not standalone parameters but computed by feeding the twist $\vec{q}$ into a multilayer perceptron with softplus activation. Please refer to the source code for more details~\cite{github}.

\section{Finite-size analysis}
As mentioned in Sec.~\ref{sec:tabc} above, twist-averaged boundary condition is a way to alleviate the finite-size error originating from the single-electron momentum shell effect. In practice, we evaluate the variational free energy and other observables by averaging over certain Monkhorst-Pack mesh (e.g., $4 \times 4 \times 4$) of the twist angle $\vec{q}$.

\begin{figure}[H]
    \centering
        \includegraphics[width=\columnwidth]{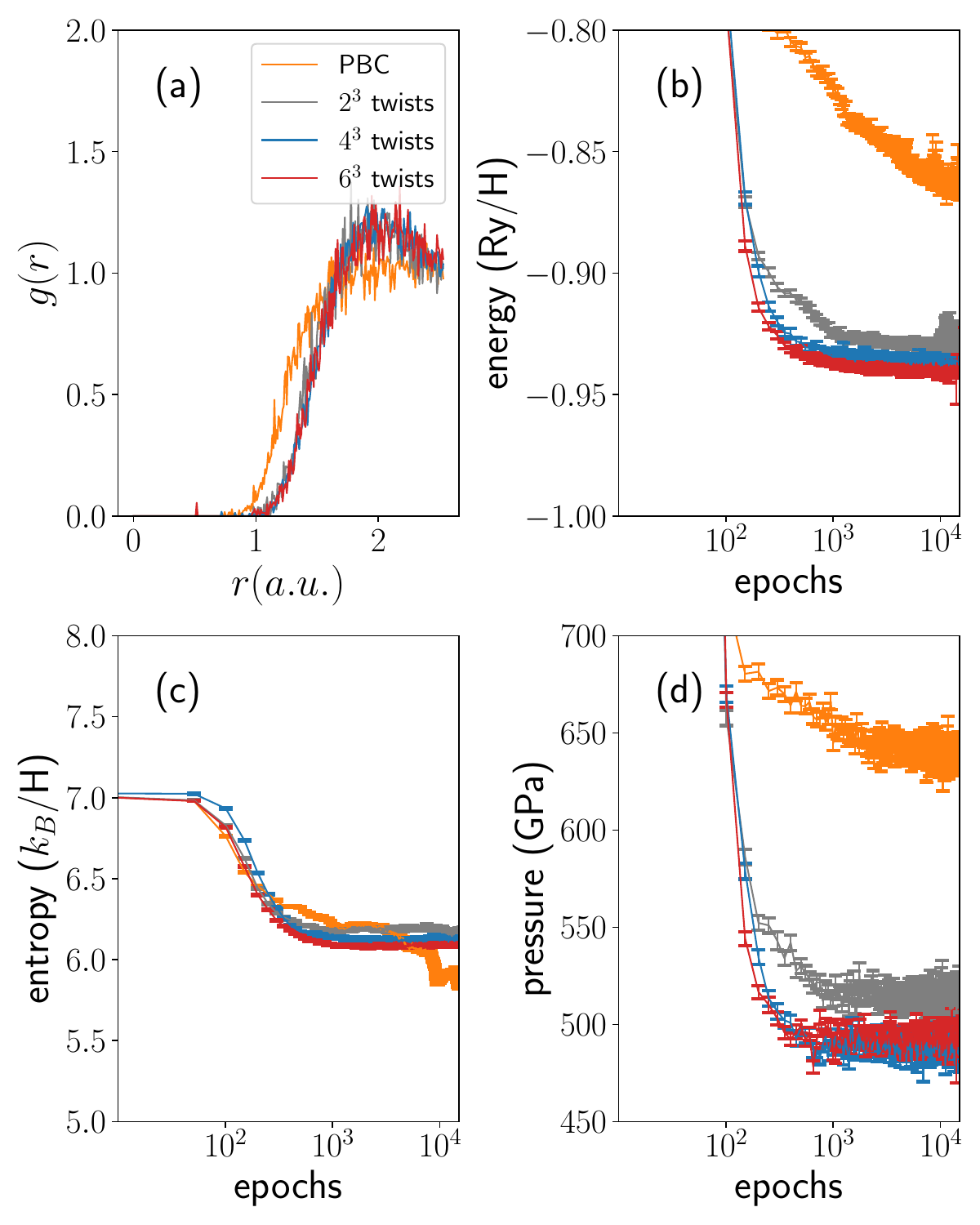}
        \caption{(a) Proton-proton radial distribution function,  (b) energy per atom,  (c) entropy per atom, and (d) pressure calculated for $N=16$ hydrogen atoms with PBC and TABC of various number of twists. The system parameters are $r_s = 1.25$, $T=6000$K.
        \label{fig:twist}
        }
\end{figure}

Figure~\ref{fig:twist} shows the proton-proton radial distribution function and thermodynamic quantities calculated for $N=16$ hydrogen atoms and various number of twists. The system parameters ($r_s = 1.25$, $T=6000$K) are the same as in the main text. One can see the twist average indeed greatly reduces the finite-size error in such a small system compared to the case of periodic boundary conditions (PBC). On the other hand, a too large twist mesh size is clearly unnecessary due to the fast convergence shown in Fig.~\ref{fig:twist}. This fact can also be justified from another point of view: due to the presence of many-body correlations, the finite-size error cannot be fully eliminated even with an infinite number of twists.

\begin{figure}[h]
    \centering
        \includegraphics[width=\columnwidth]{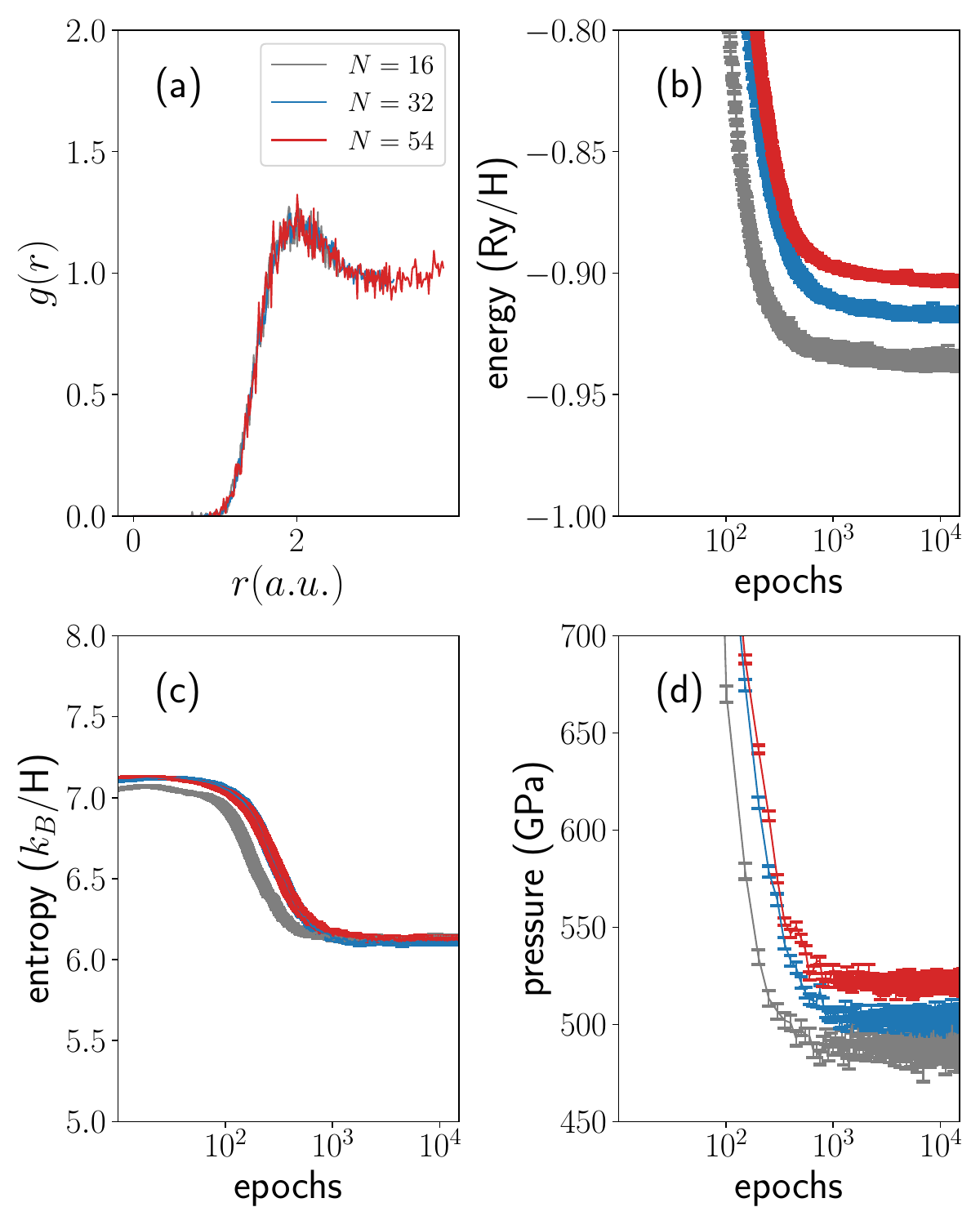}
        \caption{(a) Proton-proton radial distribution function,  (b) energy per atom,  (c) entropy per atom, and (d) pressure calculated for $N=16, 32$ and $54$ hydrogen atoms and $4^3$ twists. The system parameters are $r_s = 1.25$, $T=6000$K.
        \label{fig:size}
        }
\end{figure}

To approach the thermodynamic limit, Figure~\ref{fig:size} shows the proton-proton radial distribution function and thermodynamic quantities calculated for various system sizes $N=16, 32$ and $54$. We choose the twist mesh size to be $4^3$, which is inline with Refs.~\cite{Morales2010a, Pierleoni2016} and sufficient according to the discussion above. One can see the proton-proton radial distribution function and entropy (per atom) converge nicely, whereas the energy and pressure have a clear dependence on the system size. This is because the correlation function and entropy are sensitive only to relative energy differences, which are not affected by the finite size effect in the leading order. Note a similar observation regarding the fast convergence of proton correlation functions was also made in Fig.~5 of Ref.~\cite{Morales2010a}.

Table~\ref{tab:extrapolate} collects the converged values of thermodynamic quantities shown in Fig.~\ref{fig:size}. For completeness, we also include estimates for the thermodynamic limit by a simple $1/N$ size extrapolation, as similarly done in Ref.~\cite{Morales2010a}.

\begin{table}[h]\caption{Finite-size and extrapolated values of the energy $e$ per atom, entropy $s$ per atom, and pressure $p$ for the hydrogen system at $r_s=1.25$, $T=6000$K. \label{tab:extrapolate}}
\begin{tabular}{@{}rrrr@{}}
		\toprule
        $N$ & $e$ & $s$ & $p$  \\
        \midrule
        16 & -0.9365(8) & 6.130(6) & 489(4) \\
        32 & -0.9169(5) & 6.116(4) & 503(3) \\
        54 & -0.9038(5) & 6.125(3) & 521(2) \\
        \midrule
        $\infty$ & -0.8921(7) & 6.117(5) & 529(4) \\
        \bottomrule
\end{tabular}
\end{table}

\section{Optimization details}
As shown in the main text, the variational free energy is expressed as a two-fold expectation over the proton and electron coordinates $\vec{S}$ and $\vec{R}$ as follows:
\begin{equation}
    F = \mathop{\mathbb{E}}_{\vec{S}\sim p(\vec{S})} \left[ k_B T \ln p({\vec{S}}) + \mathop{\mathbb{E}}_{\vec{R} \sim |\psi_{\vec{S}}( \vec{R})|^2} \left[ E_{\vec{S}}^\textrm{loc}(\vec{R}) \right] \right],
\end{equation}
where $E_{\vec{S}}^\textrm{loc}(\vec{R}) \equiv \frac{H \psi_{\vec{S}}(\vec{R})}{\psi_{\vec{S}}(\vec{R})}$ is the local energy. Denoting the parameters in the proton Boltzmann distribution $p(\vec{S})$ and electron wave function $\psi_{\vec{S}}(\vec{R})$ as $\vec{\phi}$ and $\vec{\theta}$, respectively, the corresponding gradients can be written as
\begin{subequations} \label{eq:gradient}
\begin{align}
    \nabla_{\vec{\phi}} F & = \mathop{\mathbb{E}}_{\vec{S}\sim p(\vec{S})} \left[ \nabla_{\vec{\phi}} \ln p({\vec{S}}) \left( k_B T \ln p({\vec{S}}) + \mathop{\mathbb{E}}_{\vec{R} \sim |\psi_{\vec{S}}( \vec{R})|^2} \left[ E_{\vec{S}}^\textrm{loc}(\vec{R}) \right] \right) \right], \\
      \nabla_{\vec{\theta}}  F & = 2 \Re \mathop{\mathbb{E}}_{\vec{S} \sim p(\vec{S})} \left[ \mathop{\mathbb{E}}_{\vec{R} \sim |\psi_{\vec{S}}( \vec{R})|^2} \left[ \nabla_{\vec{\theta}} \ln \psi^{\ast}_{\vec{S}}( \vec{R})   \cdot  E_{\vec{S}}^\textrm{loc}(\vec{R})  \right ] \right.\nonumber \\ 
     & - \left .\mathop{\mathbb{E}}_{\vec{R} \sim |\psi_{\vec{S}}( \vec{R})|^2}  \left [  \nabla_{\vec{\theta}} \ln \psi^{\ast}_{\vec{S}}( \vec{R}) \right] \mathop{\mathbb{E}}_{\vec{R} \sim |\psi_{\vec{S}}( \vec{R})|^2} \left[ E_{\vec{S}}^\textrm{loc}(\vec{R}) \right] \right]. \label{eq:grad-theta}
\end{align}
\end{subequations}
Note that different from Ref.~\cite{Xie2022}, the ground-state wave function ansatz $\psi_{\vec{S}}(\vec{R})$ in the present work is not necessarily normalized. This results in the extra second term in the gradient estimator \Eq{eq:grad-theta}, which would be vanishing for normalized wave functions. To estimate this term accurately in practice, we sample $W$ independent proton configurations from $p(\vec{S})$, then given each of them we sample a \emph{minibatch} of electron configurations of size $B$ from $|\psi_{\vec{S}}(\vec{R})|^2$. For the calculation in the main text, we set proton batch size $W=1024$ and electron minibatch size $B=8$, respectively.
  
We jointly optimize the proton distribution $p(\vec{S})$ and electron wave function $\psi_{\vec{S}}(\vec{R})$ using the second-order stochastic reconfiguration method for variational density matrices developed in~\cite{Xie2022}. The update rules for the variational parameters read
\begin{subequations}
    \begin{align}
        \Delta\boldsymbol{\phi} &= - (\mathcal{I} + \eta \mathds{1})^{-1} \nabla_{\boldsymbol{\phi}} F, \\
        \Delta\boldsymbol{\theta} &= - (\mathcal{J} + \eta \mathds{1})^{-1} \nabla_{\boldsymbol{\theta}} F,
    \end{align}
    \label{eq:update}
\end{subequations}
where
\begin{equation}
    \mathcal{I}_{ij} = \mathop{\mathbb{E}}_{\vec{S} \sim p(\vec{S})} \left[ \frac{\partial \ln p(\vec{S})}{\partial \phi_i} \frac{\partial \ln p(\vec{S})}{\partial \phi_j} \right]
    \label{eq: Iij Bures}
\end{equation}    
is the classical Fisher information matrix, and
\begin{widetext}
\begin{align}
    \mathcal{J}_{ij} &= \Re \int d\vec{S} p(\vec{S}) \Braket{\vec{S}, \frac{\partial \psi_{\vec{S}}}{\partial \theta_i} | \vec{S}, \frac{\partial \psi_{\vec{S}}}{\partial \theta_j}} - \Re\iint d\vec{S} d\vec{S}^\prime \frac{2 p(\vec{S}) p(\vec{S}^\prime)}{p(\vec{S}) + p(\vec{S}^\prime)} \Braket{\vec{S}^\prime, \frac{\partial \psi_{\vec{S}^\prime}}{\partial \theta_i} | \vec{S}, \psi_{\vec{S}}} \Braket{\vec{S}, \psi_{\vec{S}} | \vec{S}^\prime, \frac{\partial \psi_{\vec{S}^\prime}}{\partial \theta_j}} \nonumber \\
    &= \Re  \int d{\vec{S}} p(\vec{S}) \left( \Braket{\frac{\partial \psi_{\vec{S}}}{\partial \theta_i} | \frac{\partial \psi_{\vec{S}}}{\partial \theta_j}} - \Braket{\frac{\partial \psi_{\vec{S}}}{\partial \theta_i} | \psi_{\vec{S}}} \Braket{\psi_{\vec{S} }| \frac{\partial \psi_{\vec{S}}}{\partial \theta_j}} \right) \nonumber \\ 
    &= \Re\ \mathop{\mathbb{E}}_{\boldsymbol{S} \sim p(\boldsymbol{S})} \left[ \mathop{\mathbb{E}}_{\boldsymbol{R} \sim |\psi_{\vec{S}}(\boldsymbol{R})|^2} \left[ \frac{\partial \ln \psi_{\vec{S}}^\ast(\boldsymbol{R})}{\partial \theta_i} \frac{\partial \ln \psi_{\vec{S}}(\boldsymbol{R})}{\partial \theta_j} \right] - \mathop{\mathbb{E}}_{\boldsymbol{R} \sim |\psi_{\vec{S}}(\boldsymbol{R})|^2} \left[ \frac{\partial \ln\psi_{\vec{S}}^\ast(\boldsymbol{R})}{\partial \theta_i} \right] \mathop{\mathbb{E}}_{\boldsymbol{R} \sim |\psi_{\vec{S}}(\boldsymbol{R})|^2} \left[ \frac{\partial \ln\psi_{\vec{S}}(\boldsymbol{R})}{\partial \theta_j} \right] \right]
    \label{eq: Jij Bures}
\end{align} 
\end{widetext}
is the average of the quantum Fisher information matrix over proton configurations. Note only the diagonal terms survive in the double integration term of \Eq{eq: Jij Bures}, which originates from the orthogonality of the eigenstates $\Ket{\vec{S}}$ for different proton configurations. We have added a small shift $\eta = 10^{-3}$ to the diagonal of the Fisher information matrices in \Eq{eq:update} for numerical stability. Furthermore, the norms of updates are constrained within a threshold of $10^{-3}$~\cite{Pfau2020c}.

Algorithm~\ref{alg:optimization} summarizes the implementation of a single optimization step described in this section.

\begin{algorithm}[H]
	\begin{algorithmic}[1]
		\caption{An optimization step}
        \Require Proton Boltzmann distribution $p(\boldsymbol{S})$ and electron wave function $\psi_{\vec{S}}(\boldsymbol{R})$.
        \Ensure Parameter updates $\Delta\vec{\phi}, \Delta\vec{\theta}$.
	    \State  Sample proton configurations $\vec{S}\sim p(\boldsymbol{S})$ \Comment $\mathbb{R}^{W\times N \times 3}$
        \State  Sample electron configurations $\vec{R} \sim |\psi_{\vec{S}}( \boldsymbol{R})|^2$ \Comment $\mathbb{R}^{W\times B\times N \times 3}$
	    \State  Estimate the gradients \Eq{eq:gradient}
        \State  Estimate the Fisher information matrices Eqs. (\ref{eq: Iij Bures}) and (\ref{eq: Jij Bures})
	    \State  \Return Parameter updates according to \Eq{eq:update}
		\label{alg:optimization}
	\end{algorithmic}
\end{algorithm}

\section{Illustrative calculation of the molecular phase \label{sec:molecular phase}}
Although we have been focusing on the metallic liquid state of dense hydrogen throughout the main text, in this section, we demonstrate that the present variational free energy approach can also be used to reproduce the molecular state at low densities. This clearly illustrates the potential of the new computational framework to reach a unified and reliable description for the whole phase diagram of dense hydrogen. 

\begin{figure}[h]
    \centering
        \includegraphics[width=\columnwidth]{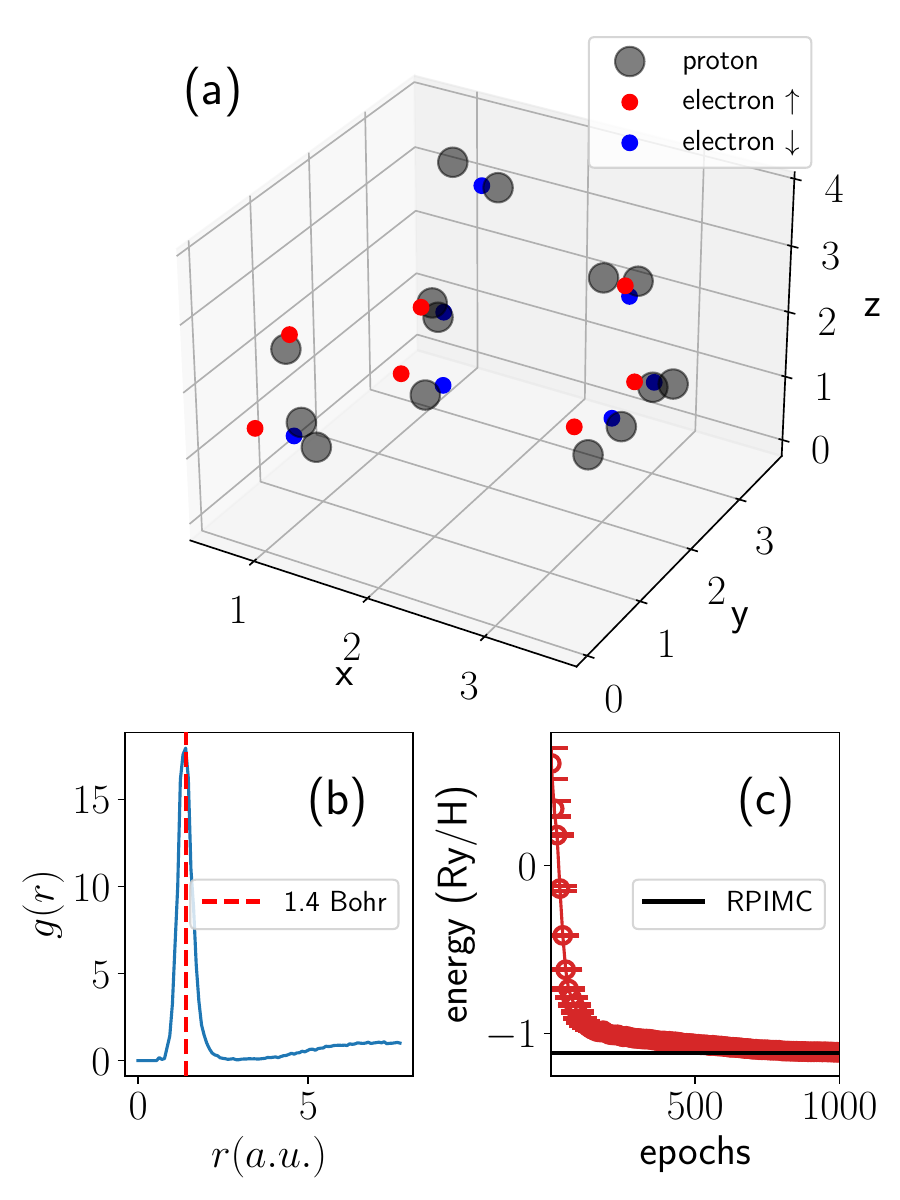}
        \caption{Deep variational free energy calculation results for $14$ pairs of protons and electrons at $r_s=4.0$, $T = 5000$K: (a) a typical snapshot of the sampled configurations. Dark dots represent protons, while red and blue dots represent spin-up and spin-down electrons, respectively; (b) proton-proton radial distribution function; (c) energy per atom as a function of optimization epochs. The black line indicates the restricted path integral Monte Carlo data from Ref.~\cite{PhysRevE.63.066404}. 
        \label{fig:mole}
        }
\end{figure}

To this end, we perform calculation for $14$ pairs of protons and electrons at $r_s=4.0$, $T=5000$K. Note the other computational setups (neural network architectures, optimization algorithm, etc.) are exactly the same as in the main text. Figure~\ref{fig:mole}(a) shows a typical snapshot of sampled configurations, which clearly contains hydrogen molecules formed by bound proton and electron pairs. The proton-proton radial distribution function shown in Fig.~\ref{fig:mole}(b) also exhibits a significant molecular peak around 1.4 Bohr; this is qualitatively different from the atomic liquid phase with itinerant electrons, as shown in \Fig{fig:epochs} of the main text. Moreover, Fig.~\ref{fig:mole}(c) shows the energy per atom as a function of optimization epochs. This result can be compared quantitatively with previous restricted path integral Monte Carlo (RPIMC) data in Ref.~\cite{PhysRevE.63.066404}.

\end{document}